\documentclass[aps,pra,onecolumn,groupaddress,amsfonts,amssymb]{revtex4}
\usepackage{amsmath}
\usepackage{amsthm}
\usepackage{xspace,amsfonts,amssymb,color}

\begin{document}

\newcommand{\bra}[1]{\mbox{$\langle\, #1|$}}
\newcommand{\ket}[1]{{|#1\,\rangle}}
\newcommand{\spg}{S{\mbox{p}}}
\newcommand{\bmu}{{\mbox{\boldmath$\mu$}}}
\newcommand{\bnu}{{\mbox{\boldmath$\nu$}}}
\newtheorem{theorem}{Theorem}[section]
\newtheorem{postulate}{Postulate}[section]
\newtheorem{principle}{Principle}[section]
\newtheorem{definition}{Definition}[section]
\newtheorem{example}{Example}[section]

  \title{The Quantum Separability Problem for Gaussian States} 
  
  \author{Stefano
    Mancini}
    \email{stefano.mancini@unicam.it}
  \affiliation{Physics Department, University of Camerino,\\
I-62032 Camerino, Italy} 
\author{Simone Severini}
\email{ss54@york.ac.uk}
  \address{Department of Mathematics, University of York,\\
YO10 5DD York, UK} 


\begin{abstract}
Determining whether a quantum state is separable or entangled is a problem
of fundamental importance in quantum information science. This is a brief
review in which we consider the problem for states in infinite dimensional
Hilbert spaces. We show how the problem becomes tractable for a class of
Gaussian states.\\
\\
\emph{Keywords:} Quantum Entanglement, Quantum Algorithms, Quantum Complexity.
\end{abstract}

\maketitle

\section{Introduction}
\label{intro}

The concept of \emph{entanglement} arose with the question of completeness
of quantum theory \cite{EPR}. Nowadays entanglement is regarded as a
fundamental property of certain quantum states and it appears to be an
important physical resource. In some sense, entanglement is synonymous of
inseparability because \emph{entangled states} possess some global
properties that cannot be explained in terms of only the parties
(subsystems) of the system. Roughly speaking, entangled states possess
\textquotedblleft strong\textquotedblright\ correlations among parties that
cannot be explained within any classical local theory (because these would
imply an instantaneous action at distance). \emph{Separable states} may also
exhibit correlations among parties, but these are purely classical and
local, hence \textquotedblleft weaker\textquotedblright\ than those
underlying entanglement.

Recently, the role of entanglement became important and often necessary in
many different contexts like quantum algorithms, quantum communication
protocols, quantum cryptography, \emph{etc.} (see \emph{e.g.} \cite{NC00}).
So, the problem of deciding whether a given quantum state is separable or
entangled has become of uppermost importance. This can be called the \emph{%
Quantum Separability Problem} (QSP). Essentially, it represents an instance
of a combinatorial optimization problems called the Weak Membership Problem 
\cite{Gro88}.

Although there exists a number of characterizations of separability, there
is still no feasible procedure to solve QSP in its generality (see \emph{e.g.%
} \cite{Ioa05} and references therein). Concerning its computational
complexity, QSP is a \textquotedblleft difficult\textquotedblright\ problem.
In fact, QSP has been proved to be NP-hard \cite{Gur03}. However, if we
restrict ourselves to specific classes of quantum states, there are examples
in which QSP can be efficiently solved. For instance, this is the case of
states in Hilbert space of dimension $2$ or $3$ \cite{PPT} and certain
finite sets of states \cite{Sev05}.

In infinite dimensional Hilbert spaces, \emph{Gaussian states} give rise to
an important class of states for which QSP is \textquotedblleft
easy\textquotedblright\ (see \emph{e.g. }\cite{j,Bra04} and the reference
therein). In this paper, we review the formulation of QSP for infinite
dimensional Hilbert spaces and we show how to tackle the problem for the
class of Gaussian states.

The paper is organized as follow. In Sec.\ref{basic} we review some basic
notions of Quantum Theory. In Sec. \ref{QSproblem} we formalize the QSP. In
Sec.\ref{Gstates} we introduce the Gaussian states. In Sec.\ref{Gcriterion}
we develop a criterion for separability of Gaussian states. Finally, some
conclusions are drawn in Sec.\ref{conclu}.

\section{Basic Notions of Quantum Theory}

\label{basic}

In this section, we introduce some terms and notions of Quantum Mechanics
needed to approach the paper. Of course, the expert reader may skip this
section.

In its standard formulation, Quantum Theory takes place in Hilbert spaces 
\cite{Dir}. A Hilbert space $\mathcal{H}$ is a vector space over the field
of complex numbers $\mathbb{C}$ endowed with an inner product (which induces
a norm), that can have finite or infinite dimension. We use the so-called
Dirac notation for a vector ${|\psi \,\rangle }$. Its dual is $%
\mbox{$\langle\,
\psi|$}$. Then, the inner product between two states ${|\psi \,\rangle }$
and ${|\phi \,\rangle }$ reads $\mbox{$\langle\, \psi|$}\phi \,\rangle \in {%
\ \mathbb{C}}$. The norm of a vector ${|\psi \,\rangle }$ results $\Vert {\
|\psi \,\rangle }\Vert =\sqrt{\langle \psi |\psi \rangle }$. The following
two postulate fix the mathematical representation of quantum states:

\begin{postulate}
The space of states of a physical system is a Hilbert space. The states are
described by unit norm vectors in such Hilbert space.
\end{postulate}

\begin{postulate}
The space of states of a composite system is the tensor product of Hilbert
spaces of subsystems.
\end{postulate}

The structure of Hilbert space naturally leads, when considering composite
systems, to the concept of \emph{entanglement}. In fact, there exist states
of the whole system that cannot be factorized into states of the subsystems.

\begin{example}
Let ${|\psi \rangle }_{1},{|\psi _{\perp }\rangle }_{1}$ be two orthogonal
states in $\mathcal{H}_{1}$ and ${|\eta \rangle }_{2},{|\eta _{\perp
}\rangle }_{2}$ be two orthogonal states in $\mathcal{H}_{2}$. Then, ${|\psi
\rangle }_{1}\otimes {|\eta \rangle }_{2}\in \mathcal{H}_{1}\otimes \mathcal{%
\ H}_{2}$ as well as $(a{|\psi \rangle }_{1}\otimes {|\eta \rangle }_{2}+b{\
|\psi _{\perp }\rangle }_{1}\otimes {|\eta _{\perp }\rangle }_{2})\in 
\mathcal{H}_{1}\otimes \mathcal{H}_{2}$, with $a,b\in \mathbb{C}$. The first
can be factorized into states of the subsystems; this is not the case for
the second one.
\end{example}

It is fashinating that this seemingly abstract mathematical notion has a
large impact in the description of the quantum mechanical world.

The above postulates can be generalized in terms of mixture of states, $%
\{p_{j},{|\psi _{j}\rangle }\}$, where $p_{j}$ denotes for the probability
for the system to be in the state ${|\psi _{j}\rangle }$. This can be done
by introducing the notion of \emph{density operator}:

\begin{definition}
A density operator $\hat{\rho}$ is a non-negative, self-adjoint, trace-one
class operator which is also positive semi-definite (that is $%
\mbox{$\langle\,
\psi|$}\hat{\rho}{|\psi \,\rangle }\geq 0$ $\forall {|\psi \,\rangle }\in 
\mathcal{H}$.
\end{definition}

Thus we can represent the mixture $\{p_{j},{|\psi _{j}\rangle }\}$ by the
density operator $\hat{\rho}=\sum_{j}p_{j}{|\psi _{j}\,\rangle }%
\mbox{$\langle\, \psi_j|$}$.

\begin{definition}
A state $\hat{\rho}$ of a composite \emph{bipartite} system is said to be 
\emph{separable} iff it can be written in the form 
\begin{equation}
\hat{\rho}=\sum_{j}p_{j}\hat{\rho}_{j}^{(1)}\otimes \hat{\rho}_{j}^{(2)},
\label{rhosep}
\end{equation}%
with non-negative $p_{j}$'s such that $\sum_{j}p_{j}=1$, and where $\hat{\rho%
}_{j}^{(1)}$, $\hat{\rho}_{j}^{(2)}$ are density operators of the
subsystems; the state is said to be \emph{entangled} otherwise.
\end{definition}

The physical quantities of a system that can (in principle) be measured are
called \emph{observables}. The next postulate fixes the mathematical
representation of observables:

\begin{postulate}
To physical observables correspond self-adjoint operators. The possible
measurement results on the observable $O$ are the eigenvalues of the
associated self-adjoint operator $\hat{O}$. The expectation value is $%
\langle \hat{O}\rangle \equiv\mathrm{Tr}(\hat{O}\hat{\rho})$.
\end{postulate}

Restrictions on expectation values are imposed by the following famous
principle:

\begin{principle}[The Uncertainty Principle]
Any two observables $A$ and $B$ in $\mathcal{H}$ must satisfy, for all
quantum states, the following inequality:%
\begin{equation}
\langle (\Delta \widehat{A})^{2}\rangle \langle (\Delta \widehat{B}%
)^{2}\rangle \geq \frac{1}{4}\left\vert \langle \lbrack \widehat{A},\widehat{%
B}]\rangle \right\vert ^{2},  \label{UP}
\end{equation}%
where $\Delta \widehat{O}\equiv \widehat{O}-\langle \widehat{O}\rangle $ and 
$[\widehat{A},\widehat{B}]\equiv \widehat{A}\widehat{B}-\widehat{B}\widehat{A%
}$ is the commutator.
\end{principle}

\section{The Quantum Separability Problem}

\label{QSproblem}

In this section we introduce the Quantum Separability Problem. Let us
consider a quantum system with two parties associated to a Hilbert space $%
\mathcal{H}_{1}\otimes \mathcal{H}_{2}\cong \mathbb{C}^{M}\otimes \mathbb{C}%
^{N}$. Notice that such a Hilbert space is isomorphic to $\mathbb{R}%
^{M^{2}N^{2}}$ and it is endowed with the Euclidean inner product $(\hat{X},%
\hat{Y})\equiv \mathrm{Tr}(\hat{X}\hat{Y})$ which induces the corresponding
norm $\Vert \hat{X}\Vert \equiv \sqrt{\mathrm{tr}(\hat{X}^{2})}$ and
distance measure $\Vert \hat{X}-\hat{Y}\Vert $. Let $\mathcal{D}\subset 
\mathcal{H}_{1}\otimes \mathcal{H}_{2}$ denote the set of all density
operators. The set of bipartite separable quantum states, $\mathcal{S}%
\subset \mathcal{D}$, is defined as the convex hull of the separable pure
states $\{|\psi \rangle_1 \langle \psi |\otimes |\eta \rangle_2 \langle \eta |\}$
where $|\psi \rangle_1 $ (resp. $|\eta \rangle_2 $) is a normalized vector in $%
\mathbb{C}^{M}$ (resp. $\mathbb{C}^{N}$). An arbitrary density matrix in $%
\mathcal{D}$ is parametrized by $M^{2}N^{2}-1$ real variables. Since we deal
with continuous quantities, in defining the separability problem we cannot
allow infinite precision, so we need to introduce a precision parameter $%
\delta \in \mathbb{R}_{+}$.

\begin{definition}[The Quantum Separability Problem]
\label{QSPdef}Given $\hat{\rho}\in \mathcal{D}$ and a precision $\delta $
assert either $\hat{\rho}$ is:

\begin{itemize}
\item \emph{Separable}: there exists a separable state $\hat{\sigma}$ such
that $\Vert \hat{\rho}-\hat{\sigma}\Vert <\frac{1}{\delta }$

or

\item \emph{Entangled:} there exists an entangled state $\hat{\tau}$ such
that $\Vert \hat{\rho}-\hat{\tau}\Vert <\frac{1}{\delta }$.
\end{itemize}
\end{definition}

In this formulation, this problem is equivalent to an instance of a
combinatorial optimization problem called Weak Membership Problem \cite%
{Gro88}. In its complete generality, QSP has been shown to be NP-hard \cite%
{Gur03}. Thus, any devised test for separability is likely to require a
number of computational steps that increases very quickly with $M$ and $N$.
For $MN\leq 6$ the positivity under \emph{Partial Transpose} (see the next
section) represents a necessary and sufficient test \cite{PPT}. Otherwise,
there only exist sufficient `one-sided' tests for separability. In these
tests, the output of some polynomial-time computable function of $\hat{\rho}$
can indicate that this is certainly entangled or certainly separable, but
not both (see \emph{e.g.} \cite{Ioa05} and reference therein).

\section{Gaussian States}

\label{Gstates}

In this section we introduce Gaussian states. Let us now move to $%
M,N\rightarrow \infty $, thus considering two infinite dimensional Hilbert
spaces $\mathcal{H}_{1}$ and $\mathcal{H}_{2}$. In such spaces we can
introduce continuous spectrum self-adjoint operators corresponding to
canonical position and momentum variables \cite{Dir}. Let us arrange them
into four-dimensional column vectors 
\begin{align}
\hat{\mathbf{v}}^{T}=(\hat{q}_{1},\hat{p}_{1},\hat{q}_{2},\hat{p}_{2}),\quad 
\mathbf{z}^{T}=(x_{1},y_{1},x_{2},y_{2}).\nonumber
\end{align}
The operators in $\hat{\mathbf{v}}$ obey commutation relations \cite{Dir}
that take the compact form 
\begin{equation}
\lbrack \hat{\mathbf{v}}_{\alpha },\hat{\mathbf{v}}_{\beta }]=i\,\Omega
_{\alpha \beta },~~~\alpha ,\beta =1,2,3,4,  \label{vv}
\end{equation}%
with 
\begin{equation}
\Omega =\left( 
\begin{array}{cc}
J & 0 \\ 
0 & J%
\end{array}%
\right) ,\;\;\;J=\left( 
\begin{array}{cc}
0 & 1 \\ 
-1 & 0%
\end{array}%
\right) .  \label{OmJ}
\end{equation}

There is a one-to-one correspondence between density operators and $c$%
-number Wigner distribution functions in phase space \cite{Wig}, the space
of variables $\mathbf{z}$, \emph{i.e.} $\mathbb{R}^{4}$, in this case.

\begin{definition}
For a given density operator $\hat\rho$ in $\mathcal{H}_1$ and $\mathcal{H}%
_2 $ the corresponding Wigner function is defined as follows\footnote{%
Throughout the paper, if not specified, the integration is intended from $
-\infty$ to $+\infty$.} 
\begin{equation}
W(\mathbf{z}):=\mathrm{Tr}\left(\hat\rho \hat{T}(\mathbf{z})\right),
\end{equation}
where 
\begin{equation}
\hat{T}(\mathbf{z}):=\frac{1}{(2\pi)^4}\int d^4\mathbf{z^{\prime}}\, \exp %
\left[i\mathbf{z^{\prime}}^T\cdot(\hat{\mathbf{v}}-\mathbf{z})\right].
\end{equation}
\end{definition}

In turn, it results 
\begin{equation}
\hat\rho=\int d^4\mathbf{z} \, W(\mathbf{z}) \hat{T}(\mathbf{z}).
\end{equation}

A density operator $\hat{\rho}$ has finite second order moments if $\mathrm{%
Tr}(\hat{\rho}\hat{q}_{j}^{2})<\infty $ and $\mathrm{Tr}(\hat{\rho}\hat{p}%
_{j}^{2})<\infty $ for all $j$. In this case we can define the vector mean $%
\mathbf{m}$ as 
\begin{align}
\mathbf{m}:=&\mathrm{Tr}(\hat{\rho}\hat{\mathbf{v}})\nonumber\\
=&\int d^{4}\mathbf{z}\,%
\mathbf{z}\,W(\mathbf{z}),
\end{align}%
and the real symmetric correlation matrix $V$ as 
\begin{equation}
V_{\alpha \beta }:=\frac{1}{2}\langle \{\Delta \hat{\mathbf{v}}_{\alpha
},\Delta \hat{\mathbf{v}}_{\beta }\}\rangle ,\quad \alpha ,\beta =1,2,3,4,
\end{equation}%
where $\{\Delta \hat{\mathbf{v}}_{\alpha },\Delta \hat{\mathbf{v}}_{\beta
}\}\equiv \Delta \hat{\mathbf{v}}_{\alpha }\Delta \hat{\mathbf{v}}_{\beta
}+\Delta \hat{\mathbf{v}}_{\beta }\Delta \hat{\mathbf{v}}_{\alpha }$ is the
anticommutator. It results 
\begin{align}
V_{\alpha \beta } &:=\mathrm{Tr}\left( \hat{\rho}\,\frac{1}{2}\{\Delta \hat{%
\mathbf{v}}_{\alpha },\Delta \hat{\mathbf{v}}_{\beta }\}\right)   \nonumber \\
&=\int d^{4}\mathbf{z}\,(\mathbf{z}-\mathbf{m})_{\alpha }\,(\mathbf{z}-%
\mathbf{m})_{\beta }\,W(\mathbf{z}).
\end{align}%
A given $V$ is the correlation matrix a \emph{physical} state iff it
satisfies 
\begin{equation}
K\equiv V+\frac{i}{2}\Omega \geq 0,  \label{Vcond}
\end{equation}%
as consequence of the Uncertainty Principle \ref{UP} and commutation
relation (\ref{vv}). The correlation matrix forms a $4\times 4$ matrix that
transforms as an irreducible second rank tensor under the linear canonical
(symplectic) transformations and has $4$ invariants. If we write the
correlation matrix in the block form 
\begin{equation}
V=\left( 
\begin{array}{cc}
A & C \\ 
C^{T} & B%
\end{array}%
\right) ,  \label{Vblock}
\end{equation}%
the invariants are $\det A$, $\det B$, $\det C$ and $\mathrm{Tr}%
(AJCJBJC^{T}J)$. The condition (\ref{Vcond}) implies $A\geq 1/4$ and $B\geq
1/4$. Moreover, Eq.(\ref{Vcond}) can be read as 
\begin{equation}
\det A\det B-\mathrm{Tr}(AJCJBJC^{T}J)-\frac{1}{4}(\det A+\det B)+\left( 
\frac{1}{4}-\det C\right) ^{2}\geq 0.  \label{unccond}
\end{equation}

It is also worth remarking that any correlation matrix can be brought into
the standard form%
\begin{equation}
V=\left( 
\begin{array}{cccc}
a &  & c &  \\ 
& a &  & d \\ 
c &  & b &  \\ 
& d &  & b%
\end{array}%
\right) ,  \label{sform}
\end{equation}%
with $a,b,c,d\in \mathbb{R}$, by effecting suitable local canonical
transformations corresponding to some element of Sp$(2,\mathbb{R})\times $ Sp%
$(2,\mathbb{R})\subset $ Sp$(4,\mathbb{R})$. Now we are ready to give the
definition of Gaussian state:

\begin{definition}
A state $\hat{\rho}$ is called \emph{Gaussian} if its Wigner function takes
the form 
\begin{equation}
W(\mathbf{z})=\frac{1}{4\pi ^{2}\sqrt{\det V}}\exp \left[ -\frac{1}{2}(%
\mathbf{z}-\mathbf{m})^{T}V^{-1}(\mathbf{z}-\mathbf{m})\right] ,
\label{PhiGauss}
\end{equation}%
with $\mathbf{m}$ a real $4$-vector and $V$ a real symmetric $4\times 4$
-matrix.
\end{definition}

One can show that $\mathbf{m}$ is indeed the mean and $V$ is the correlation
matrix. These define the Gaussian state uniquely. In what follows, we simply
consider the case $\mathbf{m}=\mathbf{0}$, because $\mathbf{m}$ can be
easily removed by some local displacement and thus has no influence on the
separability or inseparability of the state.


\section{A separability criterion for Gaussian states}

\label{Gcriterion}

In this section, we describe how to solve QSP for Gaussian states. Let us
consider a separable state ${\hat{\rho}}_{sep}$ of the form (\ref{rhosep})
in the Hilbert space $\mathcal{H}_{1}\otimes \mathcal{H}_{2}$. Let us choose
a generic couple of observables for each subsystem, say ${\hat{r}}_{j},{\hat{%
s}}_{j}$ on $\mathcal{H}_{j}$ ($j=1,2$), with 
\begin{equation}
{\hat{\mathcal{C}}}_{j}=i\left[ {\hat{r}}_{j},{\hat{s}}_{j}\right] \,,\quad
j=1,2\,.
\end{equation}%
Then, we introduce the following observables on $\mathcal{H}_{1}\otimes 
\mathcal{H}_{2}$: 
\begin{align}
{\hat{u}} &=a_{1}{\hat{r}}_{1}+a_{2}{\hat{r}}_{2}\,,  \nonumber \\
{\hat{v}} &=b_{1}{\hat{s}}_{1}+b_{2}{\hat{s}}_{2}\,,  \label{comb}
\end{align}%
with $a_{j},b_{j}\in \mathbb{R}$. From the the Uncertainty Principle \ref{UP}%
, it follows that every state ${\hat{\rho}}$ on $\mathcal{H}_{1}\otimes 
\mathcal{H}_{2}$ must satisfy 
\begin{equation}
\langle (\Delta {\hat{u}})^{2}\rangle \langle (\Delta {\hat{v}})^{2}\rangle
\geq \frac{|a_{1}b_{1}\langle {\hat{\mathcal{C}}}_{1}\rangle
+a_{2}b_{2}\langle {\hat{\mathcal{C}}}_{2}\rangle |^{2}}{4}\,.  \label{heis}
\end{equation}%
However, for separable states, a stronger bound exists. We have in fact the
following theorem \cite{Man03}:

\begin{theorem}
\label{theonec} For any separable state the following implication holds: 
\begin{equation}
{\hat{\rho}}_{sep}\quad \Longrightarrow \quad \langle (\Delta {\hat{u}}%
)^{2}\rangle \langle (\Delta {\hat{v}})^{2}\rangle \geq {\mathcal{W}^{2}}\,,
\label{th51}
\end{equation}%
where 
\begin{equation}
\mathcal{W}=\frac{1}{2}\big(\;|a_{1}b_{1}|\;\mathcal{W}_{1}+|a_{2}b_{2}|\;%
\mathcal{W}_{2}\;\big)\,,
\end{equation}%
with 
\begin{equation}
\mathcal{W}_{j}\equiv \sum_{k}p_{k}\;|\langle {\hat{\mathcal{C}}}_{j}\rangle
_{k}|\,,\qquad j=1,2,
\end{equation}%
being $\langle {\hat{\mathcal{C}}}_{j}\rangle _{k}\equiv \mathrm{Tr}[{\ \hat{%
\mathcal{C}}}_{j}{\hat{\rho}}_{k}^{(j)}]$.
\end{theorem}

The theorem can be proved with the help of a family of linear inequalities 
\begin{equation}
\alpha \langle (\Delta {\hat{u}})^{2}\rangle +\beta \langle (\Delta {\hat{v}}%
)^{2}\rangle \geq 2\sqrt{\alpha \beta }\;\mathcal{W},\qquad \alpha ,\beta
\in \mathbb{R}_{+},  \label{sumvar}
\end{equation}%
which must be always satisfied by separable states. The convolution of such
relations gives the condition (\ref{th51}), representable by a region in the 
$\langle (\Delta \hat{u})^{2}\rangle $, $\langle (\Delta \hat{v})^{2}\rangle 
$ plane delimited by an hyperbola.

Notice that, since 
\begin{equation}
{\mathcal{W}}_{j}=\sum_{k}p_{k}|\langle {\hat{\mathcal{C}}}_{j}\rangle
_{k}|\geq \left\vert \sum_{k}p_{k}\langle {\hat{\mathcal{C}}}_{j}\rangle
_{k}\right\vert =|\langle {\hat{\mathcal{C}}}_{j}\rangle |,
\end{equation}%
the following inequalities hold 
\begin{align}
\mathcal{W} &\geq \frac{1}{2}\big(\;|a_{1}b_{1}|\;|\langle {\hat{\mathcal{C}%
}}_{1}\rangle |+|a_{2}b_{2}|\;|\langle {\hat{\mathcal{C}}}_{2}\rangle |\;%
\big)  \nonumber \\
&\geq \frac{1}{2}\big(\;|a_{1}b_{1}\;\langle {\hat{\mathcal{C}}}_{1}\rangle
+a_{2}b_{2}\;\langle {\hat{\mathcal{C}}}_{2}\rangle |\;\big)\,.
\label{chain}
\end{align}%
In particular, Eq. (\ref{chain}) tells us that the bound (\ref{th51}) for
separable states is much stronger than Eq. (\ref{heis}) for generic states.
Moreover, Eq. (\ref{chain}) gives us a simple separability criterion. In
fact, while $\mathcal{W}$ is not easy to evaluate directly, as it depends on
the type of convex decomposition (\ref{rhosep}) that one is considering, the
right hand side of Eq. (\ref{chain}) is easily measurable, as it depends on
the expectation value of the observables ${\hat{\mathcal{C}}}_{j}$. Then, we
can claim that Eq.(\ref{th51}) is a necessary criterion for separability, 
\emph{i.e.} 
\begin{equation}
\langle (\Delta {\hat{u}})^{2}\rangle \langle (\Delta {\hat{v}})^{2}\rangle <%
\mathcal{W}^{2}\quad \Longrightarrow \quad {\hat{\rho}}\quad \mathrm{%
entangled}\,.
\end{equation}

\begin{example}
An important simplification applies when the observable $\hat{\mathcal{C}}%
_{j}$ is proportional to the identity operator, \emph{e.g.} $\hat{r}%
_{j}\equiv \hat{q}_{j}$ and $\hat{s}_{j}\equiv \hat{p}_{j}$. In such a case,
Eq.(\ref{heis}) reduces to 
\begin{equation}
\langle (\Delta {\hat{u}})^{2}\rangle \langle (\Delta {\hat{v}})^{2}\rangle
\geq \frac{1}{4}\,,
\end{equation}%
while Eq.(\ref{th51}) reduces to 
\begin{equation}
\langle (\Delta {\hat{u}})^{2}\rangle \langle (\Delta {\hat{v}})^{2}\rangle
\geq 1\,,
\end{equation}
\end{example}


Let us now consider the case in which $\hat{r}_{j}$, $\hat{s}_{j}$ are
linear combinations of canonical observables $\hat{q}_{j}$ and $\hat{p}_{j}$%
, \emph{i.e.} 
\begin{align}
\hat{r}_{1}\equiv \hat{q}_{1}+\frac{a_{3}}{a_{1}}\hat{p}_{1} &\quad \quad %
\hat{s}_{1}\equiv \hat{p}_{1}+\frac{b_{3}}{b_{1}}\hat{q}_{1}  \nonumber \\
\hat{r}_{2}\equiv \hat{q}_{2}+\frac{a_{4}}{a_{2}}\hat{p}_{2} &\quad \quad %
\hat{s}_{2}\equiv \hat{p}_{2}+\frac{b_{4}}{b_{2}}\hat{q}_{2}\;,
\label{newqp}
\end{align}%
where $a_{3},a_{4},b_{3}$, $b_{4}\in \mathbb{R}$ are generic real
parameters. Then Eq. (\ref{th51}), taking into account Eq.(\ref{vv}),
becomes 
\begin{equation}
\langle (\Delta u)^{2}\rangle \langle (\Delta v)^{2}\rangle \geq \frac{1}{4}%
\big(\;|a_{1}b_{1}-a_{3}b_{3}|+|a_{2}b_{2}-a_{4}b_{4}|\;\big)^{2},
\label{prod}
\end{equation}%
that should be compared with 
\begin{equation}
\langle (\Delta u)^{2}\rangle +\langle (\Delta v)^{2}\rangle \geq
|a_{1}b_{1}-a_{3}b_{3}|+|a_{2}b_{2}-a_{4}b_{4}|\;.  \label{sum}
\end{equation}%
It is easy to verify that, given $a_{j},b_{j}$ ($j=1,\ldots ,4$), the
\textquotedblleft product condition\textquotedblright\ (\ref{prod}) implies
the \textquotedblleft sum condition\textquotedblright\ (\ref{sum}). However,
if we require Eqs. (\ref{prod}) and (\ref{sum}) to be verified for \emph{all}
possible values of the coefficients $a_{j},b_{j}$, the two are equivalent
since it is possible to re-obtain one from another using a convolution trick,
like the used with Eqs.(\ref{th51}) and (\ref{sumvar}) (the one-to-one correspondence between quadratic and linear tests under all circumstances has been also pointed out in Ref.\cite{Jens}).

It turns out that the restriction 
\begin{equation}
\langle (\Delta u)^{2}\rangle +\langle (\Delta v)^{2}\rangle \geq
|a_{1}b_{1}-a_{3}b_{3}|+|a_{2}b_{2}-a_{4}b_{4}|,\quad \forall a_{j},b_{j}\in 
\mathbb{R},  \label{sim}
\end{equation}%
is \emph{necessary} and \emph{sufficient} for separability of Gaussian
states \cite{Sim00,Dua00}.

However, solving QSP by testing the condition (\ref{sim}) would be hard from
a complexity point of view, due to the presence of the universal quantifier
at right hand side.

Nevertheless, the condition (\ref{sim}) can be rephrased in a simpler way.
First notice that the uncertainty relation satisfied by all (separable and
inseparable) states 
\begin{equation}
\langle (\Delta u)^{2}\rangle \langle (\Delta v)^{2}\rangle \geq \frac{1}{4}%
|a_{1}b_{1}-a_{3}b_{3}+a_{2}b_{2}-a_{4}b_{4}|^{2},  \label{UPprod}
\end{equation}%
and corresponding to $\hat{u}$ and $\hat{v}$ formed from Eq.(\ref{th51}), is
equivalent to 
\begin{equation}
\langle (\Delta u)^{2}\rangle +\langle (\Delta v)^{2}\rangle \geq
|a_{1}b_{1}-a_{3}b_{3}+a_{2}b_{2}-a_{4}b_{4}|,  \label{UPsum}
\end{equation}%
as much as like Eqs.(\ref{prod}) and (\ref{sum}).

Then, what is the relation between conditions (\ref{sum}) and (\ref{UPsum})?

\noindent They are simply related by the \emph{partial transpose} transform 
\begin{equation}
PT:\;\;\;\hat{\mathbf{v}}\longrightarrow \Lambda \hat{\mathbf{v}}%
,\;\;\;\Lambda =\mbox{diag}(1,1,1,-1).  \label{PT}
\end{equation}%
This operation inverts $\hat{p}_{2}$, leaving $\hat{q}_{1}$, $\hat{p}_{1}$,
and $\hat{q}_{2}$ unchanged\footnote{%
The partial transposition of a density matrix, \emph{i.e.} the transposition
with respect to the subsystem $2$, is equivalent to a mirror reflection in
the phase space subsystem $2$. That is, $\hat{\rho}\rightarrow \hat{\rho}%
^{T_{2}}\Leftrightarrow W(x_{1},y_{1},x_{2},y_{2})\rightarrow
W(x_{1},y_{1},x_{2},-y_{2})$.}. In fact, separable states satisfies the
usual uncertainty relation (\ref{UPsum}) and the analogous one obtained
under partial transpose; thus these satisfy the condition (\ref{sum}).

On the other hand, the transformation (\ref{PT}) changes the correlation
matrix as $V\rightarrow \tilde{V}=\Lambda V\Lambda $. Hence, the compact
uncertainty relation (\ref{Vcond}) becomes 
\begin{equation}
\tilde{V}+\frac{i}{2}\,\Omega \geq 0.  \label{sepcond}
\end{equation}%
Expressed in terms of invariants, the condition (\ref{sepcond}) for $\tilde{V%
}$ takes a form identical to (\ref{unccond}). The signature in front of $%
\det C$ in the second term on the left hand side is changed. Thus, if we
write 
\begin{align}
f(V):= &\det A\det B+\left( {\frac{1}{4}}-|\det C|\right) ^{2}  \nonumber \\
&-\mbox{tr}(AJCJBJC^{T}J)-\frac{1}{4}(\det A+\det B),  \label{fdef}
\end{align}%
the requirement that the correlation matrix of a separable state has to obey
(\ref{sepcond}), in addition to the fundamental uncertainty principle (\ref%
{Vcond}), can be stated as follow

\begin{theorem}
\label{theof0} A bipartite Gaussian state is separable iff $f(V)\ge 0$.
\end{theorem}

The necessity follows from theorem \ref{theonec}. The sufficiency follows
from the fact that Gaussian states with correlation matrix having $\det
C\geq 0$ are separable \cite{Sim00}.

The statement \ref{theof0} is equivalent to the condition (\ref{sim}), but
much more effective to be used.

Given the standard form (\ref{sform}) of the correlation matrix $V$ we can
consider the space of all possible Gaussian states as isomorphic to $\mathbb{%
R}^{4}$, while the set of \emph{physical} states is a subspace $\mathcal{%
G\subset }$ $\mathbb{R}^{4}$ defined through (\ref{Vcond}). Furthermore, the
equation $f(V)=0$ reads 
\begin{equation}
f(V)=4(ab-c^{2})(ab-d^{2})-(a^{2}+b^{2})-2|cd|-\frac{1}{4}=0.  \label{fsf}
\end{equation}%
The equation defines the surface $S$ of the subset \textrm{$\mathcal{S}%
\subset \mathcal{G}$ }of separable states. Then, by simply evaluating $f$ we
can say whether a given state (point in \textrm{$\mathcal{G}$}) is within 
\textrm{$\mathcal{S}$} (hence separable) or not (hence entangled). This is
an easy computational task that can be efficiently accomplished. In reality,
taking into account a finite accuracy $\delta $, we can only say that the
state is \textrm{\emph{almost}} separable (resp. \textrm{\emph{almost}}
entangled) within $\delta $. Nevertheless, if we want to assert that the
state is \textrm{\emph{strictly}} separable (resp. \textrm{\emph{strictly}}
entangled), we have to be sure that the distance of the state $\hat{\rho}$
from the surface $S$ is greater than $1/\delta $. That is 
\begin{equation}
\min_{\hat{\rho}^{\prime }\in S}\Vert \hat{\rho}-\hat{\rho}^{\prime }\Vert >%
\frac{1}{\delta }.  \label{d}
\end{equation}%
According to Sec.\ref{QSproblem}, the distance between two states is
considered as $\Vert \hat{\rho}-\hat{\rho}^{\prime }\Vert \equiv \sqrt{{\rm Tr}%
\left[ (\hat{\rho}-\hat{\rho}^{\prime })^{2}\right] }$ and for Gaussian
states this can be expressed through Wigner functions (hence correlation
matrices) as 
\begin{equation}
\Vert \hat{\rho}-\hat{\rho}^{\prime }\Vert =\int d^{4}\mathbf{z}\left[ W(%
\mathbf{z})-W^{\prime }(\mathbf{z})\right] ^{2}.  \label{dPhi}
\end{equation}%
Such a task can be efficiently accomplished with the aid of geometrical
arguments and simple algorithms.
For instance, a software package that efficiently find all hyperplanes tangent to the surface $S$, from which evaluate the l.h.s. of Eq(\ref{d}), is already available \cite{Wit}.


\section{Conclusions}

\label{conclu}

Summarizing, we have given a brief review of QSP for Gaussian states of two
parties. The problem has been approached by developing tests that involve variances to 
arrive at an efficient solution based
on the invariance (positivity) of only separable states under partial
transpose. Notice that this argument can be further generalized to partial
scaling transforms to which partial transpose belongs. In fact, while $K$
and $V$ are always invariant under linear canonical transformations, they
are not invariant under scale changes on the ${\hat{\mathbf{v}}}$ that are
not contained in Sp($4,\mathbb{R}$). In particular under partial scaling $K$
is not necessarily positive definite \cite{OVM05}. These arguments could be
extended to multipartite systems, with \emph{e.g.} $N$ degrees of freedom.
Starting from the uncertainty relation $K\equiv V+\frac{i}{2}\Omega \geq 0$,
we can perform an arbitrary scaling described by the real vector $\mathbf{x}%
=(x_{1},\,x_{2},\,\ldots x_{2N})$ and then compute 
\begin{equation}
K^{\mathbf{x}}=V^{\mathbf{x}}+\frac{i}{2}\Omega ,\quad V^{\mathbf{x}%
}=\Lambda _{\mathbf{x}}V\Lambda _{\mathbf{x}},  \label{Kx}
\end{equation}%
with $\Lambda _{\mathbf{x}}\equiv {\mbox{diag}}(x_{1},\,x_{2},\,\ldots
x_{2N})$. The $2N$ real quantities $\mathbf{x}$ parameterize the Abelian
scaling semigroup with the requirement that 
\begin{equation}
|x_{1}x_{2}|\geq 1,\;|x_{3}x_{4}|\geq 1,\ldots ,|x_{2N-1}x_{2N}|\geq 1.
\end{equation}%
The necessary condition for the separability of the state is 
\begin{equation}
K^{\mathbf{x}}\geq 0,\quad \forall \mathbf{x}.  \label{Kx0}
\end{equation}%
Notice, however, that for multipartite systems besides separability (resp.
inseparability) there can be the possibility of partial separability (resp.
partial inseparability), \emph{e.g.} separability of a subsystem with
respect to the others which in turns are entangled \cite{Bra04}. Hence QSP
becomes much more subtle and even for Gaussian states it is not completely
understood.

\section*{Acknowledgemts}
We warmly thank Jens Eisert for useful comments.


\end{document}